\documentclass[11pt,letterpaper]{article}  
\usepackage{osajnl} %% do not use with REVTeX4
\usepackage[draft]{hyperref} %% optional

\begin{document}

\title{Development of Multi-Fourier Transform interferometer :Fundamental}
\author{Izumi S. Ohta}
\affiliation{National Astronomical Observatory 2-21-1 Osawa Mitaka Tokyo
181-8588, Japan\footnote{When this research was performed, Izumi
S. Ohta was with Tohoku University Aoba Aramaki Aoba-ku Miyagi 
980-8488,Japan. Izumi S. Ohta is now with National Astronomical Observatory 
2-21-1 Osawa Mitaka Tokyo 181-8588, Japan}}
%\affiliation will also work
\email{izumi.ohta@nao.ac.jp}

\author{Makoto Hattori}
\address{Tohoku University Aoba Aramaki Aoba-ku Miyagi 980-8488, Japan} %\affiliation will also work

\author{Hiroshi Matsuo}
\address{National Astronomical Observatory 2-21-1 Osawa Mitaka Tokyo 181-8588, Japan} %\affiliation will also work

\begin{abstract}
We propose the development of an instrument by the Martin \&
 Puplett-type Fourier Transform Spectrometer to applying the aperture
 synthesis technique in millimeter and submillimeter waves. We call this
 equipment the Multi-Fourier Transform interferometer(MuFT). 
MuFT performs a wide band imaging, spectroscopy and polarimetry in
 millimeter and submillimeter wavelengths. We describe the fundamentals
 of MuFT, and give an example of one potential implementation. Full
 description of the observables by MuFT are provided. A physical
 explanation of the observability of the complex visibility by MuFT is
 given. Fundamental restrictions on observations with MuFT, eg. limits
 on spectral and spatial resolutions and field-of-view, are
 discussed. The advantages of MuFT are also summarized. 
\end{abstract}

% REPLACE WITH CORRECT OCIS CODES FOR YOUR ARTICLE
% NOTE: \ocis{} IS ALIASED TO \pacs{} BUT MUST
% FORMAT THE TERMS CORRECTLY FOR EACH JOURNAL

\maketitle %% null function with osajnl.sty

\section{Introduction}

Astronomical observation of submillimeter waves represents one of the
last unexplored regions of the electromagnetic spectrum. One of the
difficulties with performing observations in this band is owing to a
large atmospheric absorption. Since submillimeter waves are a boundary 
of the radio and infrared, there remain a lot of frontiers in the 
development of fundamental observational technology.

Radio interferometry and synthesis arrays, which are basically ensembles
of two elements interferometers, are used to make measurements of fine
angular resolution\cite{Thompson}. Interferometers have been used at
millimeter waves\cite{ishiguro} and have been considered one of the
best instruments at submillimeter wave\cite{ALMA}. By setting two 
telescopes with separation larger than the maximum construct-able
diameter of the single mirror, higher angular resolution than single
dish system is easily achieved with aperture synthesis type
interferometers. By combining many telescopes, a large effective area
can be attained to increase the sensitivity of the system. 
Interferometers are robust against atmospheric fluctuations since 
large scale atmospheric fluctuations do not cause interference 
signals and are automatically omitted by the interferometers. 

However, currently operating millimeter and submillimeter
interferometers have the following two fundamental problems 
because they use heterodyne receivers as focal plane detectors: The
receiving systems perform modulation and amplification of the source
signal by mixing with a local oscillator signal. They can measure
the phase of the incident electromagnetic wave. In the following
discussion, the space-borne observation assumes as for an ideal case. 
The noise temperature of heterodyne receivers at frequency 
$\nu$ is written as $T_{sys-het}\sim 2 ({h\nu }/{k_B})\left[ (\epsilon^2
n^{2}(\nu,T_0) +\epsilon n(\nu,T_0))^{\frac{1}{2}} +1 \right]$, 
where $n$ is the thermal Planck function with $T_0=2.73~K$ due to the cosmic
microwavebackground (CMB), $h$ and $k_B$ are Planck constant and 
Boltzmann constant, respectively, and $\epsilon$ represents the detection
efficiencies. The first term represents the photon noise due to
statistical fluctuation of the CMB intensity. The second term represents 
the quantum limit of the heterodyne receivers which comes from the 
uncertainty principle between the phase and the number of 
photons\cite{Rieke}. This equation shows that the noise temperature 
is limited by the quantum limit and is increased linearly with 
the frequency in the frequency range of $\nu >100$~GHz. Therefore,
the sensitivity of heterodyne-based interferometers is severely
limited by the quantum limit of the receivers. 
Another problem is the severely limited field-of-view (FOV) of the
interferometers, due to the difficulty of constructing a large format
detector array of heterodyne receivers \cite{Rieke}. In addition,
the possible bandwidth of the interferometer is also limited by
a bandwidth of Intermediate Frequency(IF) amplifier, since the
phase change of the modulated signal after mixing with a local
oscillator signal must be followed. %

On the other hand, direct detectors such as bolometer are also used
in millimeter and submillimeter systems except for
interferometers. Direct detectors measure the intensity of the source
signal. They cannot measure the phase of the incident electromagnetic
wave. Therefore, they have not been used as focal plane detectors for
interferometers. 
The quantum limit of a direct detector is the minimum detectable power
when any internal noise can be neglected, the quantum limit of a direct 
detector is expressed in noise temperature as $T_{qu-dir}\sim {h\nu}/(k_{B}
\Delta t \Delta \nu)$, where $\Delta t$ and $\Delta \nu$ are the
response time of the detector, which is about $1-100$~msec in the case
of bolometer, and the bandwidth of the detector, which is about
$10-100$~GHz\cite{Rieke}. The noise temperature of direct detectors at
frequency $\nu$ is written as $T_{sys-dir}\sim 2 ({h\nu }/{k_B})\left[
(\epsilon^2 n^{2}(\nu,T_0) +\epsilon n(\nu,T_0))^{\frac{1}{2}} +1/
\Delta t \Delta \nu \right]$. The scecond term is negligible compared 
with the CMB photon noise at millimeter and submillimeter waves.  
Therefore, noise temperature of direct detectors are much lower 
than heterodyne receivers in the frequency range of 
$\nu >100$~GHz. Actually, the noise power of the most sensitive
currently operating bolometers is already lower than the quantum limit
of the heterodyne in the frequency range of $\nu>100$~GHz. In Contrast
to the heterodyne, a large format millimeter and submillimeter direct
detector array is possible to construct\cite{benford, Matsuo}. 
There is essentially no limit on the bandwidth for direct 
detectors\cite{Zmuidzinas}.

The above discussion indicates that the bolometric interferometer, which
is able to use a direct detector as a focal plane detector for the
interferometer, could be one of the best ideal instruments for 
millimeter and submillimeter astronomy since it shares the advantages 
of both direct detectors and interferometer. Since direct detectors
cannot obtain the information of the phase of electromagnetic waves, 
two beams captured by two apertures must be mixed before guided into 
the detectors. There are two ways to mix beams. One is 
called Fizeau type beam mixing. The simplest example of this method 
is that two beams obtained by masking a single telescope are mixed 
at the focus of the telescope. Images of different portions of the 
sky are focused on different positions of the focal plane. 
Therefore, the extension of the FOV by using the focal plane detector 
array is straightforward in this case. However,
there is a fundamental problem in applying this method to millimeter and
submillimeter observations. We have to use many detector pixels within
a single FOV to resolve the fringe pattern inside of the FOV. 
Since the direct detector arrays in millimeter and submillimeter wave
bands are still expensive, the cost performance of applying this method to
this waveband is poor. The other method is called the Michelson-type
beam mixing. An application of the Fourier Transform Spectrometer (FTS)
to aperture synthesis interferometry has been studied as a possible 
solution of the Michelson-type beam mixing bolometric interferometer. 
The application was first independently proposed by Itoh \&
Ohtsuka(1986)\cite{Itoh} and Mariotti \& Ridway (1988)\cite{Mariotti} in
near infrared (NIR). The system was referred by these authors as
''double Fourier interferometry'',  since the signal obtained by the
system is a Fourier transformation of both spectra and intensity
distribution on the sky. Itoh \& Ohtsuka studied single pupil
interferometry. Mariotti \& Ridway studied multi-pupil interferometry for
high spatial resolution. As explained in this paper, the extension of
the FOV by putting the focal plane detector array is not
straightforward, but it is possible with some restrictions.
Rinehart et al. (2004)\cite{Rinehart} have been studying the extension
of the FOV by putting a detector array on the focal plane of the double
Fourier in laboratory experiments in an optical wave bands by using optical
CCD camera. They refer to this experiment as the Wide-field Imaging
Interferometry Testbed.

In this paper, we propose the application of a Martin \& Puplett-type
Fourier Transform spectrometer (Martin \& Puplett(1969)\cite{Ma-Pu},
hereafter MP-FT) to the aperture synthesis system in millimeter and
submillimeter waves(Ohta(2004)\cite{sen}). A wire grid polarizer (WG) is
used as a beam splitter in this system. So, the wavelength dependence of
reflectivity of WG is small and is suitable for a wide band measurement
system. By setting the two input WGs appropriately, 2D intensity
distribution of four Stokes parameters can be measured. The signal obtained
by our system is a Fourier transformation of spectra and intensity
distributions of four Stokes parameters (multiple components) on the
sky. Therefore, we refer to this system as a Multi-Fourier Transform
interferometer, abbreviated to MuFT. The abbreviation  also
contains the meaning that this instrument measures Mutual correlation of
the source signal instead of auto-correlation as in the case of usual
FTS.

The plan of this paper is as follows:  
In Section 2, fundamentals of imaging and spectroscopy by MuFT are
explained using scalar waves for simplicity. 
How complex visibility is measured by MuFT is also explained in
Sec. 2. In Sec. 3, details of the components of MuFT are introduced. A
full description of the observables by the MuFT and restrictions
intrinsic to MuFT are also provided in Sec. 3. In Sec. 4, the advantages
of the MuFT are summarized. A summary of the paper is given in Sec. 5. 
This paper presents the fundamental theory of MuFT. Details of broadband
imaging and spectroscopy experiments using MuFT are presented in a
forthcoming paper. 

\section{Fundamentals of Multi-Fourier Transform interferometer}

\subsection{Fundamentals of imaging by MuFT}
\label{sec:for}

Coordinate systems on the  source plane and the observer plane are defined in
Fig. 1. The notation of this coordinate system follows Born \&
Wolf\cite{Born}. The coordinate system on the source plane ($\xi, \eta$) is
fixed on the sky. The origin of the coordinate system on the source
plane is defined to the center of the FOV. The coordinate system on the
observer plane is defined as $X$ and $Y$ axes that are parallel to $\xi$
and $\eta$ axes, respectively. The origin of the observer coordinate is
defined as the mid-point of two apertures mounted at $P_1$ and $P_2$. The
position vector connecting to the center of the aperture mounted at $P_2$ with
the center of the aperture mounted at $P_1$ is defined as 
${\bf b}=(b_x,b_y)$, which is referred to as the baseline vector. 
No absorption, no emission and no scattering due to matter distributed
between the source and observer are assumed.

We consider the measurement of the mutual coherence function of the
waves obtained by two apertures $P_1$ and $P_2$. The source electric
fields arrived at $P_{1}$ and $P_{2}$ are defined as $E_{P1}(t)$ and
$E_{P2}(t)$, respectively. They are recombined at Q. When two beams are
recombined at Q, there is a time lag $\tau$ which is caused by the
internal light path length difference from each aperture to Q, that is
$\tau=(\ell_1-\ell_2)/ c=2 x/ c$. 
The mutual coherence function with the time lag $\tau$ is defined as 
\begin{equation}
\Gamma _{12}(\tau)\equiv 
\left\langle E_{P1}(t) E_{P2}^{*}(t+\tau )\right\rangle,
\label{eq:2-1}
\end{equation}
where $\langle\rangle$ denotes time average. The intensity of the
recombined signal is written as 
\begin{eqnarray}
I_{12}(Q,\tau)&=&I_{1}+I_{2}+2\Gamma_{12}^{(r)}(\tau), 
\label{eq:2-2}
\end{eqnarray}
where $I_1$ and $I_2$ are the intensities of the signals obtained by
each aperture and $\Gamma_{12}^{(r)}(\tau)$ is a real part of
$\Gamma_{12}(\tau)$. 
The unwanted DC components could be subtracted, and are neglected in the
following discussion. Since the sizes of the astronomical objects are
much larger than the wavelengths of the electromagnetic waves which we
are interested in, any astronomical extended source can be treated as a
collection of incoherent point sources except for a few exceptional cases. A
viewer angle of the source from the observer plane is introduced as
$\mbox{\boldmath{$\theta $}}=(\theta_x,\theta_y)=(\xi/R,\eta/R)$. 
Then, the mutual coherence function and its real part can be expressed by the specific source intensity, $I(\mbox{\boldmath{$\theta $}}, \nu)$, as
\begin{eqnarray}
\Gamma_{12}({\mathbf b},\tau)&\equiv&\Gamma_{12}(\tau)=
\int _{\Omega }  \left\{ \int I(\mbox{\boldmath{$\theta$}}, \nu )\exp\left[ -2\pi j \frac{\nu }{c}(\mathbf{b}\cdot 
\mbox{\boldmath{$\theta $}})+2\pi j \nu \tau \right] \>{\rm d}\nu \right\} {\rm d}^2\theta ,\\ 
\Gamma_{12}^{(r)}({\mathbf b},\tau)&\equiv&\Gamma_{12}^{(r)}(\tau)=
\int _{\Omega }\left\{ \int I(\mbox{\boldmath{$\theta $}}, \nu )\cos \left[-2\pi \frac{\nu }{c}({\mathbf b}
\cdot\mbox{\boldmath{$\theta $}})+2\pi \nu \tau \right]
\> {\rm d}\nu \right\} {\rm d}^{2}\theta ,
\label{eq:bx1}
\end{eqnarray}
where $\Omega $ is the solid angle expanded by the source, $j$ is an
imaginary unit, and $\Gamma_{12}({\mathbf b},\tau)$ and
$\Gamma_{12}^{(r)}({\mathbf b},\tau)$ explicitly express the dependence
of the mutual coherence function on the baseline vector. The
intensity expressed in Eq. (4) is the observable by MuFT.

The spectroscopy-resolved source image is extracted from the observed
$\Gamma_{12}^{(r)}$ in the following way:  
For each fixed combination of ${\mathbf b}$, $\Gamma_{12}^{(r)}$ is
measured for the time lag from $-\tau_0$ to $\tau_0$ by changing
$\ell_2$ continuously. By performing a Fourier integration of the
obtained $\Gamma_{12}^{(r)}$ in $\tau$, the spectrally resolved mutual
coherence function, $\hat{\Gamma}_{12}({\mathbf b},\nu)$, is obtained as 
\begin{eqnarray}
\hat{\Gamma}_{12}({\mathbf b} ,\nu )&=&\int _{\Omega }
\left\{\int I(\mbox{\boldmath{$\theta $}}, \nu') 2\tau_0 {\rm sinc} \left[ 2\pi (\nu'-\nu)\tau_0 \right]
\exp\left[ -2\pi j\frac{\nu'}{c}({\mathbf b}\cdot
     \mbox{\boldmath{$\theta $}})\>{\rm d}\nu' \right] \right\}{\rm
d}^2\theta ,
\label{eq:bnu1}
\end{eqnarray}
where $I(\mbox{\boldmath{$\theta $}}, -\nu)=I(\mbox{\boldmath{$\theta
$}}, \nu)$ is used and ${\rm sinc}\> x=(\sin x)/x$. This is a kind of
Winner-Khenchine Formula\cite{Born}. 
The integration by $\nu'$ reduces to $\nu '=\nu $ with spectral 
resolution of $\Delta \nu\sim 1/\tau_0$. Since $\nu \gg 1/\tau_0$ 
in an usual case, $\Delta\nu\ll \nu$. Now the customary notations 
in the interferometer can be introduced, that is 
\begin{equation}
u=\frac{b_{x}\nu }{c}=\frac{b_{x}}{\lambda }\, \, ,
\, \, v=\frac{b_{y}\nu }{c}=\frac{b_{y}}{\lambda }.
\label{eq:uv}
\end{equation}
The $(u, v) $ is called spatial frequency. When $| {\mathbf b}\cdot
\mbox{\boldmath{$\theta $}} |\Delta\nu/c \ll 1$, the Fourier
transformation of the mutual coherence function can be written as  
\begin{equation}
\hat{\Gamma}_{12}(u,v,\nu)=\int _{\Omega } 
\tilde{I}(\mbox{\boldmath{$\theta $}}, \nu) \exp\left[-2\pi 
j(u\theta_x+v\theta_y) \right]\> {\rm d}^2\theta ,
\label{eq:bnu2}
\end{equation}
where $\tilde{I}$ is a spectrally convolved source intensity distribution 
defined as
\begin{equation}
\tilde{I}(\mbox{\boldmath{$\theta $}},\nu)=
\int I(\mbox{\boldmath{$\theta $}}, \nu') 
2\tau_0 {\rm sinc} [2\pi (\nu'-\nu)\tau_0]\> d\nu' .
\label{eq:fts}
\end{equation}
The functional form of the convolution kernel depends on the apodization
function adopted when the Fourier transformation by $\tau$ is
performed. In the above example, the top hat function is adopted as the
apodization function. Eq. (7) is equivalent to the equation which
expresses the van Cittert-Zernike Formula in Fraunhofer
limit\cite{Born}. It shows that the Fourier transformation of the mutual
coherence function, $\hat{\Gamma}_{12}(u,v,\nu)$, is the complex
visibility function of the source except for the proportionality constant. 
The main reason why the MuFT is capable of measuring the source complex
visibility function, is to take data about the time lag $\tau$ from 
$-\tau _{o}$ to $ \tau _{o}$ symmetrically. By 
Fourier transforming them in $u$ and $v$, source images are obtained for
various frequencies. The MuFT is, in principle, is capable of
performing simultaneous measurements of source image and spectrum. 
The MuFT corresponds to the XF-type radio
interferometer\cite{Thompson}. Physical explanation of the
principles of imaging by MuFT is given in Sec.2B. 

The observability of the complex visibility function by MuFT makes
observations with MuFT easier. From Eq. (7) 
$\hat{\Gamma}_{12}^{*}(u, v,\nu)=\hat{\Gamma}_{12}(-u, -v, \nu)$. 
Here we have used the fact that the source intensity distributions, 
$I(\mbox{\boldmath{$\theta $}},\nu)$, are a real variable called the
reality condition. Therefore, sampling in only half of the $u v$ plane
is enough to perform imaging observations by MuFT. 

\paragraph{Large dynamic range advantage}
When the source intensity distribution does not depend on frequency, all
the source images obtained for various frequencies can be summed up into
a single image in the following way: In this case, the frequency
dependence $i(\nu )$ and the spatial coordinate dependence of the
source intensity $B(\mbox{\boldmath{$\theta $}})$ are decomposable, such
as $I(\mbox{\boldmath{$\theta $}}, \nu )=B(\mbox{\boldmath{$\theta
$}})i(\nu )$. The spectrally convolved source intensity distribution can
also be written as $\tilde{I}(\mbox{\boldmath{$\theta $}},\nu
)=B(\mbox{\boldmath{$\theta $}})\tilde{i}(\nu )$
where $\tilde{i}(\nu)$ is the convolution of $i(\nu)$ with the spectral
convolution kernel. Eq. (7) can be rewritten as 
\begin{eqnarray}
\hat{\Gamma}_{12}(u,v,\nu )
&=&\int_{\Omega}B(\mbox{\boldmath{$\theta $}})\> \tilde{i}(\nu)
\exp\left[-2\pi j\frac{\nu}{c}({\mathbf b}\cdot \mbox{\boldmath{$\theta $}})\right]\>{\rm d}^{2}\theta .
\label{eq:Obs-2}
\end{eqnarray}
Divided by $\tilde{i}(\nu)$, the frequency dependence of the source intensity can be removed. 

Suppose that the observations for the baseline vector combinations on a
circle with a radius of $|{\mathbf b}|=b$ on the baseline vector plane were
performed with continuous sampling. Observed frequency bandwidth is
supposed to be very wide. Adopt a polar coordinate on the baseline
vector plane as ${\mathbf b}=(b,\varphi)$. Integrate
$\hat{\Gamma}_{12}({\mathbf b},\nu ) \exp[2\pi j \nu ({\mathbf b}\cdot
\mbox{\boldmath{$\theta $}})/c ]/\tilde{i}(\nu)$ by $\nu {\rm d}\nu$ and
$(b^2/c^2 ){\rm d}\varphi$ for a fixed $b$. Transform variables from
$\nu$ and $\varphi$ to $u= b \nu\cos \varphi/ c $, $v=b \nu \sin
\varphi/ c$ using 
$\partial(u,v)/\partial(\nu,\varphi) =\nu b^2/ c^2$.
This procedure leads to the source intensity distribution as  
\begin{eqnarray}
&&\int_{0}^{2\pi}\left\{\int_{0}^{\infty}\nu  
\frac{\hat{\Gamma}_{12}({\mathbf b},\nu )}{\tilde{i}(\nu)}\exp\left[2\pi j \frac{\nu}{c}
({\mathbf b}\cdot \mbox{\boldmath{$\theta $}})\right]\>{\rm d}\nu \right\}\> \frac{b^2}{c^2}{\rm d}\varphi \nonumber \\
&=&\int_{0}^{2\pi}\left\{\int_{0}^{\infty}\nu 
\int_{\Omega}{\rm d}^2\theta' B(\mbox{\boldmath{$\theta $}}') \exp\left[-2\pi j\frac{\nu}{c}({\mathbf b}
\cdot \mbox{\boldmath{$\theta $}}')\right]\exp\left[2\pi j
\frac{\nu}{c}({\mathbf b}\cdot \mbox{\boldmath{$\theta $}})\right]\>{\rm d}\nu \right\}\> \frac{b^2}{c^2}{\rm d}\varphi ,\nonumber \\
&=&\int _{-\infty}^{\infty}\left( \int_{\Omega}B(\mbox{\boldmath{$\theta $}}')
\exp \left\{ 2\pi j\left[u(\theta'_x-\theta_x)+v(\theta'_y-\theta_y)\right]\right\} \>
{\rm d}^2\theta' \right)\>{\rm d}u{\rm d}v, \nonumber \\
&=&\int_{\Omega} B\bigl(\mbox{\boldmath{$\theta $}}')\delta ^{2}
(\mbox{\boldmath{$\theta $}}'-\mbox{\boldmath{$\theta $}})\>{\rm d}^2\theta', \nonumber \\
&=&B(\mbox{\boldmath{$\theta $}}).
\end{eqnarray}
In essence, 
the procedure can be thought of 
as follows: 
Since the spatial distribution is independent of frequency,
the choice of frequency to measure the source intensity distribution
is arbitrary. For a fixed baseline interval, observations with short
wavelengths measure precise structures with high resolution and
observations with long wavelengths measure overall structures
with low resolution. By gathering all this information,
precise source intensity distribution is obtained.

Several modifications are required to apply the above procedure 
to real observations. First of all, sampling on the baseline vector
plane can neither be perfect nor continuous. The sampled points on this
plane will be discrete and sparse. 
Instead of the Fourier integration described in Eq. (10), 
discrete Fourier transformation must be performed. 
To express the sparseness of the sampled points, 
convolution with a window function which is set at zero at the baseline
vector points where the observation was not performed, must be taken
into account in this calculation. 
Secondly, the source spectrum may not be finite for all frequencies. 
The source intensity may be zero in some frequency ranges
due to absorption by the intervening medium or no emission from the source.
Division by $\tilde{i}(\nu)$ described in Eq. (10) will diverge 
in these frequency ranges. 
The data for these frequency ranges should be removed in the procedure.
Otherwise, they may introduce a large amount of noise into the obtained map. 
These modifications result in an imperfect map. However, it does not
mean that the intrinsic power of the large dynamic range advantage is
lost. The imperfection of the synthesized map due to the imperfect $uv$
coverage is one of the most common problems in interferometer. The
large dynamic range advantage of MuFT can be explained as a large $uv$
coverage being obtained from a small number of baseline vector
combinations. This technique 
is essentially the same as the Multi-Frequency Synthesis 
(MFS) used in radio interferometers (Conway et al.(1990)\cite{conway}).

The situation to which this procedure is applicable 
is a very special case.
However, there are plenty of 
astronomically important objects whose spatial distribution is
independent on frequency, such as millimeter and submillimeter wave 
intensity distribution of clusters of galaxies known as 
Sunyaev-Zel'dovich effect (hereafter SZ effect, Sunyaev \& Zel'dovich 
(1972)\cite{SZ}, Hattori \& Okabe(2005)\cite{H-O}). 

\subsection{The physical explanation of imaging by MuFT}

As shown in the previous section, the essential reason 
why MuFT is capable of performing imaging observations in broadband is
that the complex visibility function of the source can be extracted from the 
observables with MuFT.  
This is accomplished by changing the time lag from $-\tau_0$ to $\tau_0$
symmetrically. 
The reason why the complex visibility function is measurable 
by this process, is physically explained as follows.

Consider a broadband point source observation by MuFT. 
Suppose that there is a geometrical time lag, $\tau_g$,  
caused by the light path length difference from the source to each 
aperture $P_1$ and $P_2$, that is $\tau_g\equiv (R_1-R_2)/c$. 
A burst-like interference signal is observed as scanning in $\tau$
by changing $\ell_2$ continuously at $\tau=-\tau_g$, where the 
geometrical time lag is compensated by the internal time lag. 
Since this burst-like signal always appears at $\tau=0$ in the FTS, 
the signal appearing at $\tau=0$ is called a central burst of the
interferogram. We also refer to this burst-like signal observed for 
a point source in MuFT to a central burst as following the FTS. 
In the case of MuFT, the burst center appears 
at the position where the total time lag of the light led by 
each aperture is zero. The position of the central burst observed 
by MuFT contains information about the position of the point source in
the sky. If the source position is slightly 
shifted from the origin of the source plane toward the same 
side of the aperture $P_1$ and $\tau_g$ is negative,
then $\tau$ must be positive and vice versa 
to observe the central burst signature. 
Suppose that $\tau$ is only scanned in the positive range. 
The central burst of the second source 
is not observable by this scan and we cannot know the position of the
second source. By rotating the 
baseline vector by 180 degrees if possible, the geometrical time 
lag for the second source becomes negative and the central burst caused 
by the second source becomes observable by the positive $\tau$ scan. 
On the other hand, the central burst of the first source goes out 
of the positive $\tau$ region. 
This rotation corresponds to take a $-{\mathbf b}$ configuration of two
apertures.  Therefore, when $\tau$ is scanned only in the positive range, 
to obtain a full source image, measurements with ${\mathbf b}$ and
$-{\mathbf b}$ configurations are required. However, if we allow for
scanning of the negative $\tau$ region, the central burst of the second source
can be detected in the negative $\tau$ region without rotating the
baseline vector. Therefore, by performing a scan in $\tau$ from the negative
to the positive range symmetrically, the full source image is measurable
without a performing 180-degree rotation of the baseline vector. 
This is why the symmetrical scan in $\tau$ leads us 
to obtain the complex visibility function. 

In contrast to the MuFT, the complex visibility function is obtained in
the following ways in the heterodyne system: 
The real part of the complex visibility function is obtained by 
directly correlating the signals obtained by each aperture. 
The imaginary part is obtained by correlating 
the signals after applying the $\pi/2$ phase shifter for one of the signals. 

\section{The application of the Martin \& Puplett-type FTS to aperture synthesis}
\subsection{Components of MuFT}

Fig.\ref{fig:MuFT0} is a schematic diagram of the MuFT system. 
From the top of the page, the parts depicted are the Light Collecting
unit(LiC), the Fourier Interference unit(FI) and the Detection \&
Sampling unit(DeS). The detected signal is transfered to a computer after it
has been through an electrical 
amplifier, a noise filter and an AD converter. 
The sampled data are analyzed by a data analysis system. 
Since the data is transformed into Polarimetry, 
Spectroscopy and Imaging information in the data analysis system, 
the data analysis system is referred to as PSI.

The LiC first part performs a division of the wave front 
of the incident source signal, and guides the signal to the Fourier 
Interference unit. Two different types are considered. 
One type has multi-apertures installed on separated multi-platforms.
The antenna in multi-platforms type 
radio interferometer and siderostat in multi-platforms type 
optical-IR interferometer correspond to this type. 
In this type, there are large geometrical time lags except for the moment 
when the source is at southing.  
To compensate for the geometrical time lags which are also time-variable
due to earth rotation, a large and time-variable 
offset in the scanning center of $\tau$ must be introduced in MuFT.  
To do this, a fringe tracking delay-line technique can be
applied\cite{Shao}.
When the MuFT system is applied to multi-platforms type LiC, 
performing a symmetrical scan in $\tau$ from the negative to the positive 
region is essential to obtain 
the complex visibility function. This is because the possible rotation 
angle of the baseline vector relative to the astronomical sources 
is limited, since the sources set and are not observable for  
nearly half a day except for sources around a celestial pole. 
The other type of LiC has that multi-apertures are installed on a
single platform. For simplicity, we consider the two-aperture case. 
The positions of the two telescopes are controlled by controlling the
azimuth and elevation angles of the platform so that a fixed reference
point on the targeted source is maintained by a baseline vector
bisector line, perpendicular to the baseline vector. The same reference
point must be kept in the center of the FOV of each telescope. Examples
of this type of LiC are CBI\cite{CBI} and AMiBA\cite{AMiBA} which have
multi apertures installed on a single platform. 
Two mirrors connected by one rail mounted on a turntable, such as SPIRIT
(Mather \& Leisawitz(2000)\cite{Mather1}), could be a solution.
In the space, using multi-satellites could be an attractive solution. 

The Fourier Interference unit is an instrument in which divided 
light beams are recombined after modulating one of the light path 
lengths to make an internal time lag and is the heart of the MuFT system.
The method of beam mixing is the Michelson type, as mentioned in the
introduction. This is designed by applying the MP-FT to aperture
synthesis. It has two entrance windows to receive two light beams guided
from the light collecting unit. Wire grid polarizers are used as beam
splitters and beam combiners. One example of the FI unit is shown in
Fig.\ref{fig:FI}. 

The direction of the WGs 1 and 2 wires must be set in the following ways. 
Consider light rays which enter the WGs 1 and 2 along the vertical axis. 
These rays are reflected by the WGs and travel toward the Roof Top Mirrors 
(RTM) along a horizontal axis, and are then reflected by RTMs and travel
back toward the WGs 1 and 2. Consider the projection of the WGs 1 and 2
on the planes normal to these light rays. The wires of these WGs must be
tilted 45 degrees from the vertical axis on these planes. There are two
possibilities for the wires being tilted clockwise (CW) or counter
clockwise (CCW) from the vertical axis. There are four combinations for
the alignment of the WGs 1 and 2. Two basic configurations of the WGs
are illustrated in Fig.\ref{fig:Op1}. When the wires of both WGs are set
in a CW direction, Stokes parameters of $\mathcal{I}$ and $\mathcal{Q}$
are observable. This is referred to as option 1. By setting both WGs to
CCW, complementary data for the first case is obtained. One of the
configurations gives $\mathcal{I}+\mathcal{Q}$ and the other gives
$\mathcal{I}-\mathcal{Q}$. When the wires of WG 1 are set CW and those
of WG 2 are set CCW direction, Stokes parameters of $\mathcal{U}$ and
$\mathcal{V}$ are observable. This is referred to as option 2. By
changing the setting of WGs 1 and 2 vice versa, complementary data for
the original combination are obtained. One of the configurations gives
$\mathcal{U}+\mathcal{V}$ and the other gives 
$\mathcal{U}-\mathcal{V}$. Therefore, by performing observations with
all four combinations, spectrally resolved source intensity
distributions of 4 Stokes parameters are obtained. Since only three of 4
Stokes parameters are independent for completely polarized waves,
systematic errors could be self calibrated in MuFT by measuring all 4
Stokes parameters for the completely polarized light beams. 
For the intensity measurement, observations with option 1 is enough. 
Details are explained in the next subsection.  

The other WGs installed in the FI unit for each option must be aligned
in the following ways: In option 1, the wires of BS WG must be directed
vertically or horizontally as illustrated in Fig.\ref{fig:FI} to combine
the two beams taken by WGs 1 and 2 at BS.
The combined beam, either reflected or transmitted by BS, is led to WGs A
or B. To get interference terms of the two beams, the wires of output
WGs A and B must be tilted 45 degrees from the vertical axis on the
projected plane normal to the transmission axis. In option 2, the wires
of BS WG projected on the same plane as the projection plane for WG 2
must be perpendicular to those of WG 2. Then, the beam taken by WG 1 is
transmitted and the beam taken by WG 2 is reflected by BS WG, and those
are led to output WG A. The wire of output WG A must be vertical or
horizontal to get interference terms of the two beams. 

To extract the maximum ability of MuFT, direct detectors like a
bolometer, which performs broadband intensity measurement, must be mounted
on the DeS unit. One of two bolometers detects transmitted signals
through the output WG, while the other detects reflected signals. 
The difference in reflected and transmitted signals reproduces the 
full mutual coherence function since the phase difference in the
interferograms between two signals is $\pi$. Also, unwanted DC
components are removed by this process. The DC fluctuations originating
in the atmospheric fluctuation and the thermal fluctuation of the
detector systems is mostly common to the transmitted and reflected
components. Therefore, these fluctuations are supposed to be removed by
taking the difference of each signal. This technique is also applicable
to MP-FT. 

The PSI system analyzes the acquired data. Also, this unit performs 
mechanical control of MuFT. Baseline length vector source tracking and
monitoring, mirror control, WG control and data acquisition are
performed. Data analysis is also performed by the PSI system. 
Details of data analysis are explained in a forthcoming paper.

\subsection{Detailed description of observables of MuFT}

In this section, fundamentals of MuFT are fully described by treating
electro-magnetic waves as vector waves. 
Suppose the electric field of the incident electromagnetic waves emitted from 
sky position $\mbox{\boldmath{$\theta $}}$ on the source plane at
frequency $\nu $ described by 
$\vec{E}(\mbox{\boldmath{$\theta $}}, \nu, t)=(E_1 ,
E_2)$ where $E_1$ and $E_2 $ are orthogonal components of the
electromagnetic waves. They can be written as,
\begin{eqnarray}
E_1 &=& \epsilon _1 (\mbox{\boldmath{$\theta $}}, \nu )\cos (2\pi \nu t + \delta _1
) = \epsilon _1(\mbox{\boldmath{$\theta $}}, \nu )\frac{\exp[2\pi j\nu t + j\delta _1]
+\exp[-2\pi j\nu t - j\delta _1]}{2},\nonumber \\
E_2 &=& \epsilon _2 (\mbox{\boldmath{$\theta $}}, \nu )\cos (2\pi \nu t +\delta _2
) = \epsilon _2(\mbox{\boldmath{$\theta $}}, \nu )\frac{\exp[2\pi j\nu t + j\delta _2]
+\exp[-2\pi j\nu t - j\delta _2]}{2}.
\end{eqnarray}
The phases for the fixed frequency, $\delta_{1}$ and $\delta_{2}$, can
be treated as time-independent variables. In option 1, WG1 and WG2
reflect the same polarization components of the incident electromagnetic
waves. Therefore, option 1 measures the mutual coherence of the same
component of the electromagnetic waves. In option 2, the components
reflected by WG 1 and WG 2 are orthogonal to each other. Therefore,
option 2 measures the mutual coherence of the orthogonal polarization
components of the electromagnetic waves. For each option, there are two
independent possible alignments of WGs, as explained in the previous
subsection. The observed intensity transmitted through WG A for each
option is expressed as 
\begin{eqnarray}
\label{eq:Op12-1}
I_{Op1}({\mathbf b},x)&=&\frac{1}{4}\int_\Omega \left(\int A_{\nu
}(\mbox{\boldmath{$\theta $}})\frac{{\epsilon _{1}}^{2}}{2}\left\{2 + \exp\left[ -2\pi j\nu\left( {\mathbf b}\cdot \mbox{\boldmath{$\theta $}}
 -\frac{2x}{c}\right) \right] \right. \right.\nonumber \\
&&+\left. \left. \exp\left[2\pi j\nu\left({\mathbf b}\cdot \mbox{\boldmath{$\theta $}}
			      -\frac{2x}{c}\right)\right]
   \right\}\>{\rm d}\nu \right)\>{\rm d}^2\theta , \nonumber \\
I_{Op2}({\mathbf b},x)&=&\frac{1}{2}\int_\Omega \left(\int \frac{A_{\nu
}(\mbox{\boldmath{$\theta $}})}{2}
\left\{ \epsilon _1 \epsilon _1 + \epsilon_2 \epsilon _2 \vphantom{\frac{A_{\nu
}(\mbox{\boldmath{$\theta $}})}{2}}\right. \right. \nonumber \\
&&+ \epsilon _1 \epsilon _2 
\exp[j(\delta _1 -\delta _2)] \exp\left[-2\pi j\nu\left({\mathbf b}\cdot\mbox{\boldmath{$\theta $}}-\frac{2x}{c}\right)\right]\nonumber \\
&&\left. \left. + \epsilon _1 \epsilon _2 \exp\left[-j(\delta _1
 -\delta _2)\right] \exp\left[2\pi j\nu\left({\mathbf b}\cdot \mbox{\boldmath{$\theta $}} -\frac{2x}{c}\right) \right] \right\}\> {\rm d}\nu \right)\> {\rm d}^2\theta , 
\end{eqnarray}
where time averages were performed and 
$A_{\nu}(\mbox{\boldmath{$\theta $}})$ expresses a beam pattern of the
observational system at frequency $\nu$ 
toward $\mbox{\boldmath{$\theta $}}$. The factors of $1/4$ and $1/2$
appearing in front of each integral come from the transmissivity and
reflectivity at BS WG and output WG. Stokes Parameters are defined in
the usual manner 
\begin{eqnarray}
\label{eq:stokes}
\mathcal{I}(\mbox{\boldmath{$\theta $}},\nu )&=&\left[ \epsilon_1(\mbox{\boldmath{$\theta $}},\nu )
\epsilon_1(\mbox{\boldmath{$\theta $}},\nu )
+ \epsilon_2(\mbox{\boldmath{$\theta $}},\nu ) \epsilon_2(\mbox{\boldmath{$\theta $}},\nu ) \right],\nonumber \\
\mathcal{Q}(\mbox{\boldmath{$\theta $}},\nu )&=&\left[ \epsilon_1(\mbox{\boldmath{$\theta $}},\nu ) \epsilon_1(\mbox{\boldmath{$\theta $}},\nu ) 
-\epsilon_2(\mbox{\boldmath{$\theta $}},\nu ) \epsilon_2(\mbox{\boldmath{$\theta $}},\nu ) \right], \nonumber \\
\mathcal{U}(\mbox{\boldmath{$\theta $}},\nu )&=&2\left[ \epsilon_1(\mbox{\boldmath{$\theta $}},\nu
)\epsilon_2(\mbox{\boldmath{$\theta $}},\nu )
\cos \delta \right] ,\nonumber \\
\mathcal{V}(\mbox{\boldmath{$\theta $}}, \nu )&=&2\left[ \epsilon_1(\mbox{\boldmath{$\theta $}},\nu ) \epsilon_2(\mbox{\boldmath{$\theta $}},\nu )
\sin \delta \right],
\end{eqnarray}
where $\delta\equiv \delta_2-\delta_1$ is the phase difference between 
the orthogonal components of the incident electromagnetic waves. 
By rewriting Eq.(12) in Stokes parameters, we have 
\begin{eqnarray}
\label{eq:Op12-2}
I_{Op1}&=&{1\over 4}\int_{\Omega}\left\{\int
A_{\nu}(\mbox{\boldmath{$\theta $}})
\frac{\mathcal{I}(\mbox{\boldmath{$\theta $}},\nu)\pm \mathcal{Q}(\mbox{\boldmath{$\theta $}},\nu)}{2}
\left[1+\cos2\pi \frac{\nu}{c}({\mathbf b}\cdot \mbox{\boldmath{$\theta $}}-2x)\right]\>{\rm d}\nu \right\}\>{\rm d}^2 \theta ,
\nonumber \\
I_{Op2}&=&\frac{1}{2}\int_{\Omega}\left\{\int A_{\nu}(\mbox{\boldmath{$\theta $}})
{1\over 2}\left[\mathcal{I}(\mbox{\boldmath{$\theta $}}, \nu
 )\vphantom{\frac{1}{2}}+ \left(\mathcal{U}(\mbox{\boldmath{$\theta
			   $}},\nu )\cos[2\pi \frac{\nu}{c}
 ({\mathbf b}\cdot \mbox{\boldmath{$\theta $}} -2x)]\right. \right. \right. \nonumber \\
&&\left. \left. \left.
\pm \mathcal{V}(\mbox{\boldmath{$\theta $}},\nu)\sin[2\pi \frac{\nu}{c}({\mathbf b}\cdot \mbox{\boldmath{$\theta $}} -2x)]\right) \right]\>{\rm d}\nu \right\}\> {\rm d}^2\theta , 
\end{eqnarray}
where $\pm$ corresponds to the alignment of the WGs, that is to CW or
CCW alignments described in the previous subsection. These results show
that the MuFT is capable of measuring all intensity distributions of 4
Stokes parameters in a wide band, eg. 
$\mathcal{I}(\mbox{\boldmath{$\theta $}},\nu), \mathcal{Q}(\mbox{\boldmath{$\theta $}},\nu),
\mathcal{U}(\mbox{\boldmath{$\theta $}},\nu)$ and $\mathcal{V}(\mbox{\boldmath{$\theta $}},\nu)$.
When the incident light is unpolarized, $\delta_1$ and $\delta_2$ are
uncorrelated, and therefore, option 2 output is zero. 
For unpolarized sources, the intensity distribution can be measured 
by using option 1(Ohta(2001)\cite{sen1} , Hattori et al.(2000)\cite{hattori}). 
Details of experimental results of polarimetry with MuFT are
presented in a forthcoming paper. 

\subsection{Spatial and spectral resolutions}

Spectral resolution is determined by the maximum internal time
lag, $\tau_0$, as $\Delta \nu\sim 1/\tau_0$. 
Spatial resolution for a fixed wavelength, $\lambda$, is defined in the 
same way as that of an conventional interferometer, and is 
limited by the maximum baseline length to $\lambda/B_{max}$. 
Since the wavelength coverage is very wide in the case of MuFT, 
the maximum spatial resolution is determined by the spatial resolution 
for minimum observable wavelength.

\subsection{The field of view of the MuFT}

Suppose broadband point-source observations with two apertures using
MuFT. Define the center of the field of view (FOV) as the origin of the
source coordinate. When the source position is $\vec{\theta}$, 
there is a geometrical time lag of $\tau_g={\mathbf b}\cdot
\mbox{\boldmath{$\theta $}}/c$. 
For simplicity, we assume that $\mbox{\boldmath{$\theta $}}$ 
is parallel to ${\mathbf b}$ 
and ${\mathbf b}\cdot\mbox{\boldmath{$\theta $}}=b\theta$. 
As explained in Sec.2.B, the central burst of the interferogram 
for this source appears 
at $\tau=-\tau_g$. 
To detect the central burst for this source, $\tau_0$ must be larger 
than $\tau_g$. 
This condition provides a limit on the FOV of MuFT of
$\theta < (c\tau_0) / b$.
A similar constraint on the FOV is deduced in a different way. 
As discussed in the previous subsection, 
the finiteness of the maximum internal time lag leads to the spectral
uncertainty of $\Delta \nu \sim 1/\tau_0$. The frequency uncertainty
results in a phase uncertainty of $\delta\phi = 2\pi \Delta \nu b
\theta/c$. To be able to observe a visibility function without
significant coherent loss, $\delta\phi$ should be less than $\pi/2$. 
This condition provides almost the same limit on the FOV as
shown in the above inequality. 
The limit shown above constrains the FOV only in the direction 
parallel to the baseline vector. 
In the direction perpendicular to the baseline vector, 
the above condition does not constrain the FOV. 
However, since the baseline vector must be rotated relative 
to the source plane for image synthesis, 
the same coverage of the FOV in both directions might be 
convenient. 
Therefore, the maximum available pixel number of the detector array 
installed on the focal plane of the MuFT is provided by 
$({\rm c}\tau_0/b \theta_p)^2$, where $\theta_p$ is the FOV of one detector pixel.

The analysis of the data measured by an off-axis pixel is not 
self-evident. 
Since the position of the central burst is shifted from the scanning 
center, the Fourier transformation of the obtained interferogram in $\tau$ 
from $-\tau_0$ to $\tau_0$ causes a phase shift of $2\pi\nu\tau_g/c$. 
Therefore, the obtained phase should be shifted back by $2\pi\nu\tau_g/c$.
There are other non-trivial problems related to using the multi-pixel 
detector array on MuFT which must be studied to optimize methods. 
Rinehart et al. (2004)\cite{Rinehart} have been studying 
the application of the multi-pixel detector array on 
the double Fourier in laboratory experiments in optical wave bands using
an optical CCD camera. They have successfully detected interferograms in
each pixel which are each observing different FOV. 

\subsection{Phase errors, efficiencies and noise}
There are four sources of phase errors:  
(1) sampling error due to shift in zero optical path difference;
(2) uncertainty of 
$(u,v)$ points due to finiteness of spectral resolution;
(3) uncertainty of 
$(u,v)$ points due to pointing
error. 
(4) Atmospheric phase fluctuation.
The first one can be compensated by zero-path calibration of the 
interferometer. The second one provides a limit on the FOV, as discussed
in the previous subsection. 
The third one can be corrected by pointing observations of compact
sources with a certain interval. Atmospheric phase error can be
corrected when atmospheric opacity and phase error are correlated. 

As seen in Eq.(12) and Eq.(14), optical efficiency of MuFT 
for a single output port is 1/8 for option 1 when unpolarized 
radiation is observed, combination of two output ports, such 
as that used for detecting reflection and transmission of WG A, 
would give an efficiency of 1/4. Detecting a signal traveling toward
WG B would give an efficiency of 1/2. 
Furthermore, setting another MuFT under the 
first one to detect the radiation transmitted through WGs 1 and 2 
would give an efficiency of 1.

Sources of noise in observed interferograms are atmospheric radiation,
radiation from optical components, internal detector noise and other
excess noise such as readout electronics noise.
Since most of these noises are incoherent, they do not create coherent peaks
in an interferogram but appear as random noise. Common amplitude
fluctuation of atmospheric emission could be removed by taking the
difference of the interferograms obtained by signals transmitted through
and reflected by output WG. 
Radiation from optical components causes a small amount of coherent
noise signal in an interferogram. Most of the noise emission from the 
optical components which propagate back to the input wire grids is
ejected to the sky. However, the direction of noise emission propagated
from optical components has a finite distribution, and finite errors in
alignment of wire grids and optical components are not escapable. 
Therefore, a small amount of radiation returns to the detector. 
This results in an autocorrelation noise emission signal from the
optical components in the interferogram. 
This can be removed by subtracting the background interferogram 
obtained by observations of the sky without source.
However, it is impossible to observe exactly the same sky 
as during source observations without the source. We have to 
adopt data obtained by observations of the sky without the 
source as the background data. 

\section{Advantages of MuFT}

One of the most prominent advantage of MuFT is that one 
can use direct detectors like a bolometer as an interferometer focal 
plane detector. This give MuFT the following three significant advantages. 

\paragraph{High sensitivity and wide bandwidth interferometer in FIR}

As described in the introduction, 
the sensitivity of direct detectors is higher than that of
heterodyne receivers in the frequency range of $\nu>100$~GHz, that is
$\lambda <3$~mm.
Therefore, the sensitivity of MuFT could be higher than an interferometer 
based on heterodyne receivers, especially in submillimeter and FIR wave bands. 
Performing wide band observations with direct detectors 
is easily achieved. For example, bolometers which have a sensitivity in
extremely broadband have been used as focal plane detectors of FTS in
FIR wave bands.  Since the WGs have uniform reflectivity and transmissivity 
for all frequencies in millimeter, submillimeter and FIR, the MuFT is
suited to broadband observations. 

\paragraph{Wide field of view}

Constructing large format heterodyne focal plane arrays is not yet 
feasible, since local oscillators and associated electrical circuits 
for each detector are bulky and not amenable to automated production
using integrated circuit technology. Therefore, not a single detector is
used in radio interferometers, including ALMA. This severely limits the
FOV of the interferometer. On the other hand, large format focal plane
arrays of direct detectors are already operating. New types of direct
detectors aiming at a dramatic increase in the number of pixels are
lying developed\cite{benford, Matsuo}. Since MuFT can use 
direct detectors as interferometer focal plane detectors, 
it could extend the FOV of the millimeter and submillimeter 
interferometers dramatically. 
Details are described in Sec.3.D.

\paragraph{Large dynamic range}

When source intensity distribution is independent of frequency, 
we can sum up all images obtained at different frequencies 
into a single image, as described in Sec.2.A. In other words, a single
baseline observation is equivalent to observations with many
combinations of baseline interval in the case of a single frequency
interferometer. Long wavelengths observe overall source structures with 
low spatial resolution. Short wavelengths observe fine structures with
high spatial resolution. A large dynamic range of spatial resolution is
covered even by a single baseline observation. A similar technique has
been addressed as MFS for heterodyne interferometer(Conway et
al.(1990)\cite{conway}). This is a technique for combining maps obtained
for different frequencies by different interferometers into a single
image. 

\paragraph{Linear and Circular Polarimetry}

The MuFT can measure all 4 components of Stokes parameters 
(Hattori et al.(2000)\cite{hattori}), as explained in sec. 3.B. 
This is because MuFT uses WGs as beam-splitter. Since a double Fourier
uses dielectric film as a beam-splitter, it cannot perform polarimetry.  

\section{Summary}
We proposed a new type of bolometric interferometer named MuFT by
applying MP-FT to aperture synthesis. Fundamentals of imaging,
spectroscopy and polarimetry with this instrument were developed. The
MuFT is a system which permits imaging and spectroscopy in a wide band
by combining the Wiener-Khinchine Formula\cite{Born} which makes it
possible to extract a spectrum from the auto-correlation and van Cittert
Zernike Formula\cite{Born} which allows imaging from the mutual
correlation. The concrete composition of this equipment was proposed. By
combining wire grid beam-splitters adequately, source intensity
distributions of four Stokes parameters can be acquired in a wide
band. Fundamental restrictions in practical use were discussed.

We are planning to perform the observations from good mm and submillimeter 
observation sites, such as the Nobeyama Radio Observatory in Japan, 
the Atacama desert in Chile, South Pole etc..  
The possibility of having a super-wide FOV 
by mounting the focal plane array of bolometers, 
SIS photon detectors and transition edge sensors is attractive for 
future applications. Optical designs of MuFT for 
using detector arrays is going to be advanced. 

Applying this technique to a space-borne mission is one of the best 
possibilities for extracting the maximum ability of MuFT, since 
there is no restriction on the bandwidth from atmospheric absorptions. 
Mather et al. \cite{Mather} has been proposing a space-borne 
FIR observatory based on this kind of technique. 
The future
application of this technique to observations from space 
could open new and interesting possibilities in FIR astronomy. 

%%%%%%%%%%%%%%%%%%%%%%%%%%%%%%%%%%%%%%%%%%%%%%%%%%%%%%%%%%%%%
\section*{ACKNOWLEDGMENTS}

The research was financially supported by the COE program
"Exploring New Science by Bridging Particle-Matter Hierarchy" at Tohoku
University, the Sasagawa Scientific Research Grant and 
the Japan Science Society and by the Grant-in-Aid for Scientific
Research (116204010) of the Japan Society for the Promotion of
Science. We would like to thank Ms. Y. Luo of Tohoku university and 
Ms. Y. McLean of arkitext.com for critical reading of English. 
At last, thanks are due to Dr. J. Nishikawa of NAOJ for helpful 
suggestions and comment about established optical and IR 
interferometers system.

\newpage

\section*{List of Figure Captions}

Fig. 1.Coordinate systems and designations used in observation of
extended source. Source is assumed to be very far away.

Fig. 2. Simplified schematic diagram of MuFT. Wave-front division of source light is performed by LiC. These are combined by FI, through optical
systems. Interferogram is measured by DeS.

Fig. 3. Example of FI unit. WG position is Option1. 
Thin dashed lines show light paths. Arrows attached to the dashed 
lines indicate the propagation direction of the light beams. 

Fig. 4. Two basic configurations of wire grids 1 and 2. The left figure
is Option.1 (WGs are parallel) and the right figure is Option.2 (WGs are
perpendicular). Gray lines are light paths. 
\newpage

\begin{figure}
\begin{center}
\includegraphics[width=15cm]{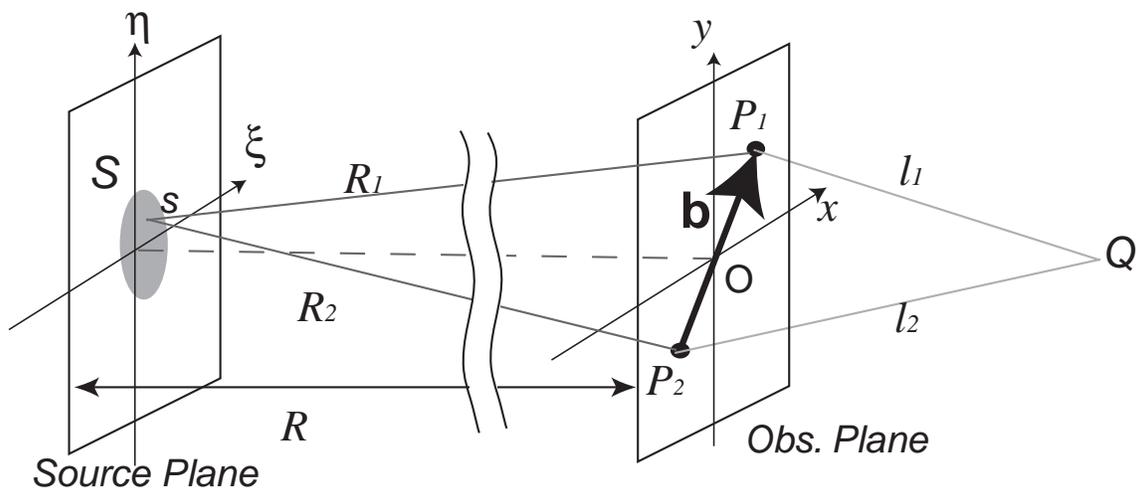}
\caption{Coordinate systems and designations used in observation of
 extended source. Source is assumed to be very far away.}
\label{so-obs} 
\end{center}
\end{figure}
\newpage

\begin{figure}[hbp]
\begin{center}
\includegraphics[width=15cm]{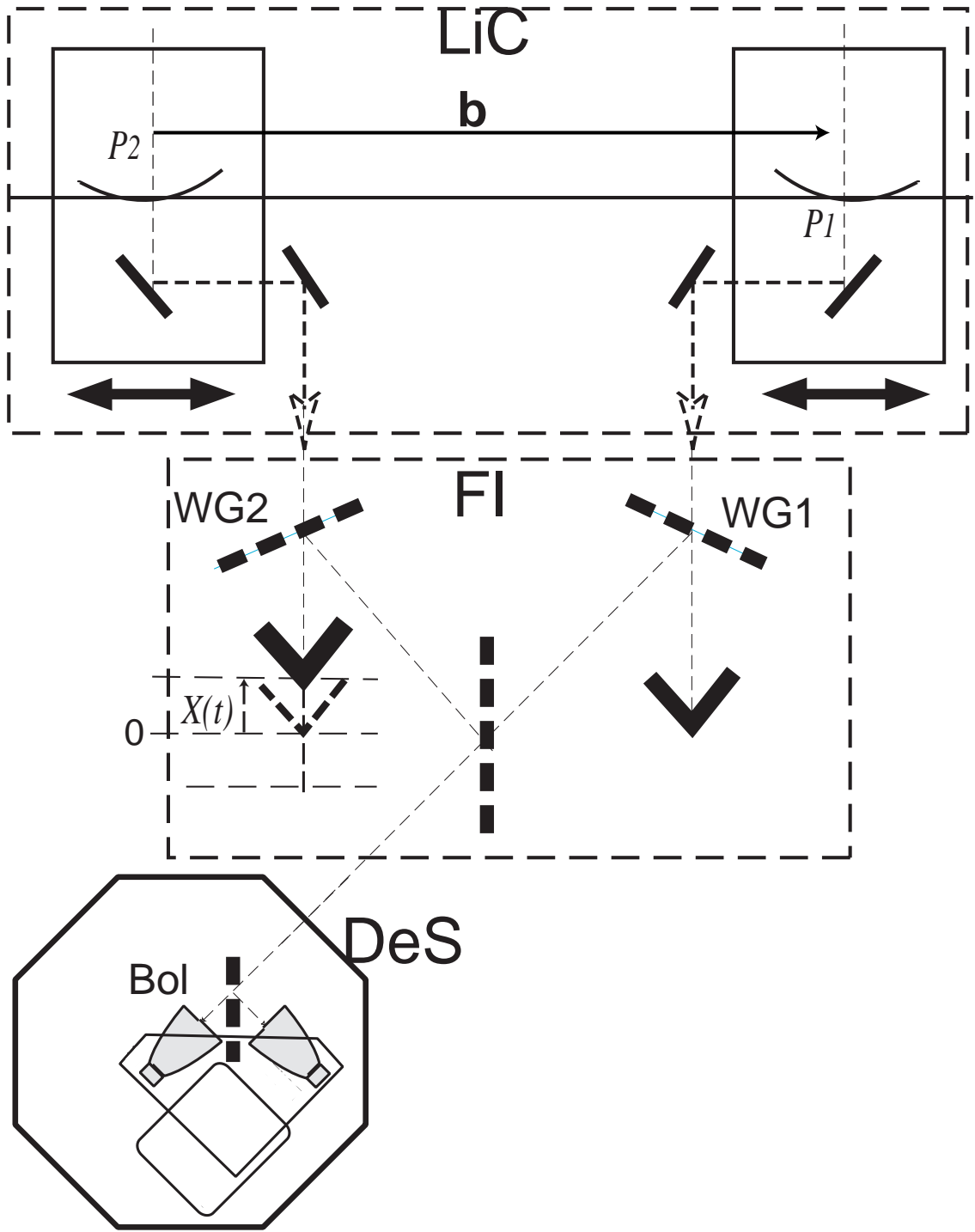}
\caption{Simplified schematic diagram of MuFT. Wave-front division of
 source light is performed by LiC. These are combined by FI, through
 optical systems. Interferogram is measured by DeS.}
\label{fig:MuFT0}
\end{center}
\end{figure}
\newpage

\begin{figure}
\begin{center}
\includegraphics[width=15cm]{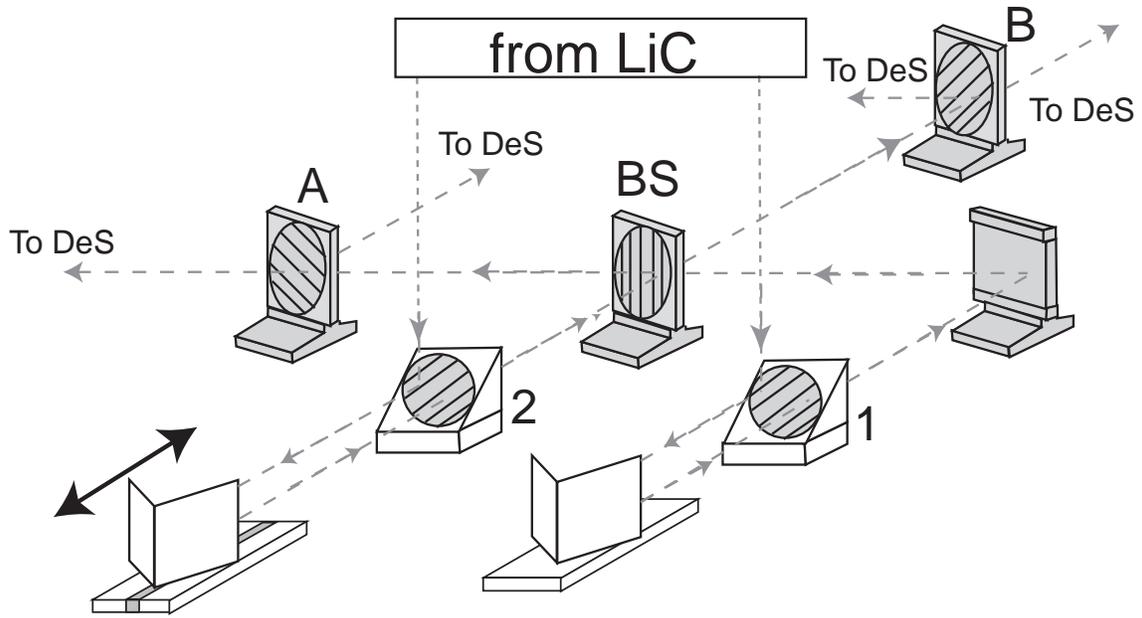}
\caption{\label{fig:FI}Example of FI unit. WG position is Option1. 
Thin dashed lines show light paths. Arrows attached to the dashed 
lines indicate the propagation direction of the light beams. }
\end{center}
\end{figure}
\newpage

\begin{figure}
\begin{center}
\includegraphics[width=15cm]{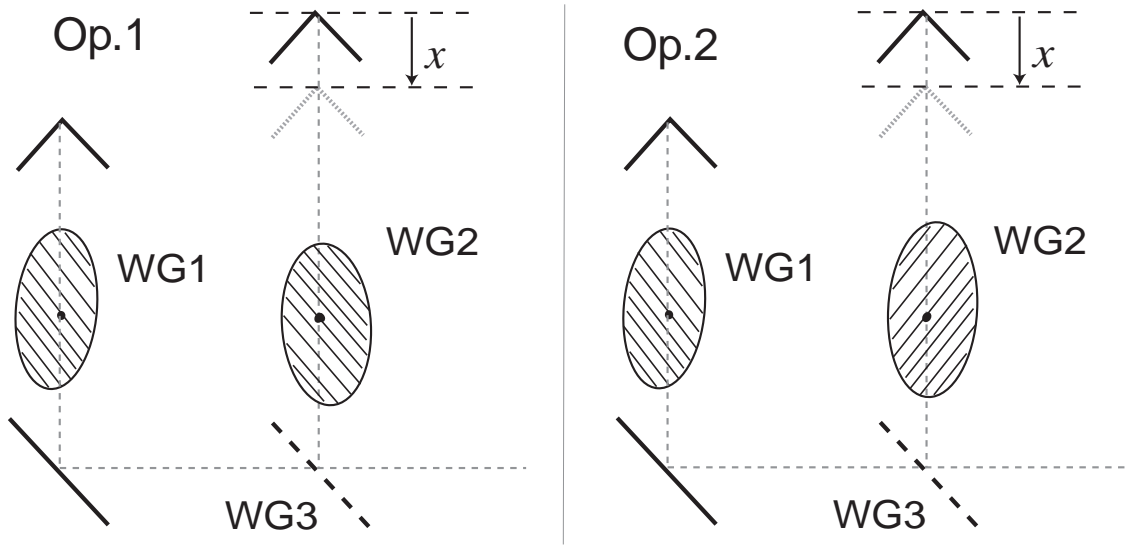}
\caption{Two basic configurations of wire grids 1 and 2. The left figure
is Option.1 (WGs are parallel) and the right figure is Option.2 (WGs are
 perpendicular). Gray lines are light paths. }
\label{fig:Op1}
\end{center}
\end{figure}

\end{document}